\documentstyle[12pt]{article}
\title{\bf T-Duality in Sigma-Models with Kaluza-Klein Metric as Electric-Magnetic Duality }
\vspace{20mm}
\author{M. A. Jafarizadeh$^{a,b}$ \thanks{E-mail: tabriz\_u@vax.ipm.ac.ir} , A. Rezaei-Aghdam$^{a,c}$\\ 
\\
\\
$^a${\small Department of Theoretical Physics and Astrophysics , Tabriz University, Tabriz  51664, Iran.} \\   
$^b${\small Institute for Studies in Theoretical Physics and Mathematics, Tehran 19395-1795, Iran.}\\ 
$^c${\small Department of Physics, Tarbiyat Moallem University, Tabriz  P.O.Box 51745-406, Iran.}}  

\begin{document}
\maketitle
\vspace{15mm}
\begin{abstract}
It is shown that the T-duality in $\sigma$-model with Kaluza-Klein metric, without 
or with a torsion term, can be interpreted as electric-magnetic duality for some of their   
solitonic solutions. Actually Buscher's duality transformation interchanges the topological  
and Noether charges.
\end{abstract} 
\newpage

\section{Introduction}
The duality symmetry in high energy physics has attracted considerable attention in recent   
years. On the one hand T-duality symmetry in string theory and $\sigma$-models inter-relates      
the theories with different geometries and shows their equivalence which leads to classification 
of string solutions and helps to know more about the vacuum states. On the other hand, the electric-magnetic    
duality in 4-dimensional supersymmetric gauge theories inter-relates the asymptotic behaviour of
strongly coupled electric theories with weakly coupled magnetic theories. Indeed this gives a new insight into the long standing quark   
confinement hypothesis, Higgs mechanism and mathematical physics. Of course, there are enough works connecting 
these two dual symmetries (for a review refer to \cite{Duff}) such that one can deduce the electric-magnetic   
dualiy from T-duality. In all these works the relation between these two symmetries has been   
studied by considering the effective action of string theory and the corresponding equations of motions (the equations of motions   
follow from vanishing of $\sigma$-model $\beta$-functions ), specially for the solitonic solutions (for more details see \cite{Khuri}).   
Therefore, all these investigations have quantum nature (first quantization), irrespective of the order of perturbation.      
Hence it is the main object of this article to investigate this connection classicaly, without using $\beta$-functions.    
We hope that this will help us to understand deeply the relation between these two dualities. To achieve this goal we investigate the   
$\sigma$-model with Kaluza-Klein metric. Of course the motion of string over Kaluza-klein space has already been studied \cite{Nielsen},    
using Nambu-Goto action for the string. Here we consider the $\sigma$-model action and assume that the world surface has an intrinsic,    
but not an embeded, metric. This assumption is made for compatibility of some special solitonic solutions which we are going to study below. Comparing  
the dual model with the original one, it has been shown that for these special solitonic solutions, T-duality can be interpreted as the electric-magnetic duality in both      
models with or without torsion terms. 
\\
The structure of the article is as follows.
In order the article to be self-contained and also to introduce the notations, first we give a brief review of abelian duality in section $2$. Then in subsection $3.1$ we consider the $\sigma$-model with Kaluza-Klein metric without torsion      
term and its equations of motion. We show that for some special solitonic solutions, these equations can reduce to geodesic equation of a charged particle. Further information      
about these solutions follow by studying the Hamiltonian of the reduced system and also by investigating the Noether and topological currents of this $\sigma$-model. In subsection $3.2$   
we obtain its dual model and study further the topological and Noether currents of the dual model. It has been shown that the T-duality is the same as the electric-magnetic duality. In subsection $4.1$     
we consider a $\sigma$-model with Kaluza-Klein metric with torsion term (Kalb-Ramond term), where we have used the non-local gauge potential of reference \cite{Cardoso}. Here also by investigating the equations 
of motion and the Hamiltonian in the case of some solitonic ansatz, we show that these solutions can be interpreted as a system describing the motion of dyon in a curved background field. The electric and magnetic charges     
of dyon have been calculated in $4.2$. Its dual model has been studied and finally it has been shown that the T-duality interchanges the electric and magnetic charge of dyon. The paper is ended with a brief conclusion.       

\section{Abelian Duality}
In this section we briefly review the Buscher's formulation of abelian duality (for more details see [5,6]). 
Consider the following two dimensional $\sigma$-model over d-dimensional target manifold $M$ with coordinate $\{X^A\}$; $A=0,1,...,d-1$ :
\begin{equation}
S=-\frac{1}{4\pi \alpha^\prime}\int\!d^2\xi[\sqrt{-h}h^{\alpha \beta}G_{AB}{\partial_\alpha}X^A{\partial_\beta}X^B+\epsilon^{\alpha \beta}B_{AB}{\partial_\alpha}{X^A}{\partial_\beta}{X^B}],
\end{equation}
where $\xi=(\xi^0, \xi^1)$ are coordinates of the world sheet and $h^{\alpha \beta}$ is its metric, while $G_{AB}$ and $B_{AB}$ are metric and torsion-potential of target manifold $M$, respectively.    
The constant $\alpha^\prime$ has dimension of length two. If $U(1)$ isometry exists, then in the adapted coordinate, isometry appears as the translation symmetry along the isometry coordinate.       
Denoting the isometry coordinate by $y$ such that ${X^A}=\{y ,x^\mu\}$ where $\mu=1,...,d-1$ , the action (1) takes the following form:      
$$ 
S=-\frac{1}{4\pi \alpha^\prime}\int\!d^2\xi[\sqrt{-h}h^{\alpha \beta}(G_{00}{\partial_\alpha}{y}{\partial_\beta}{y}+2G_{0 \mu}{\partial_\alpha}{y}{\partial_\beta}{x^\mu}+G_{\mu \nu}{\partial_\alpha}
{x^\mu}{\partial_\beta}{x^\nu})
$$
\begin{equation}
\hspace{20mm} +\epsilon^{\alpha \beta}(2B_{0 \mu}{\partial_\alpha}{y}{\partial_\beta}{x^\mu}+B_{\mu \nu}{\partial_\alpha}{x^\mu}{\partial_\beta}{x^\nu})].
\end{equation}
In order to obtain the dual model we make this isometry local by gauging it   
(i.e. replacing the derivatives of isometry by covariant derivative) so that with the help of     
lagrange multiplier, this gauge field is constrained to flat one. Then     
by simply integrating over the gauge field, having gauged away the isometry coordinate,  
we are led to the following dual model: 
$$ 
\tilde{S}=-\frac{1}{4\pi \alpha^\prime}\int\!d^2\xi[\sqrt{-h}h^{\alpha \beta}(\tilde{G}_{00}{\partial_\alpha}{\chi}{\partial_\beta}{\chi}+2\tilde{G}_{0 \mu}{\partial_\alpha}{\chi}{\partial_\beta}{x^\mu}+\tilde{G}_{\mu \nu}{\partial_\alpha}
{x^\mu}{\partial_\beta}{x^\nu})
$$
\begin{equation}
\hspace{20mm} +\epsilon^{\alpha \beta}(2\tilde{B}_{0 \mu}{\partial_\alpha}{\chi}{\partial_\beta}{x^\mu}
+\tilde{B}_{\mu \nu}{\partial_\alpha}{x^\mu}{\partial_\beta}{x^\nu})],
\end{equation}
where
$$
\tilde{G}_{00}=1/G_{00}\hspace{3mm},\hspace{3mm}\tilde{G}_{0 \mu}=B_{0 \mu}/G_{00}\hspace{3mm},\hspace{3mm}\tilde{G}_{\mu \nu}=G_{\mu \nu} -\frac{G_{0 \mu}G_{0 \nu}-B_{0 \mu}B_{0 \nu}}{G_{00}}
$$

\begin{equation}
\hspace{10mm}\tilde{B}_{0 \mu}=G_{0 \mu}/{G_{00}}\hspace{3mm} ,\hspace{3mm} \tilde{B}_{\mu \nu}=B_{\mu \nu}-\frac{G_{0 \mu}B_{0 \nu}-G_{0 \nu}B_{0 \mu}}{G_{00}}.
\end{equation}
From above relations we see that the dual model possesses $U(1)$ isometry too.   
Equations (1) and (3) have the same physical results, even        
though the target manifold of given $\sigma$-model and its dual differ geometrically. Notice that these 
transformations can be obtained throught the following canonical transformation [6,7]: 
\begin{equation}
P_y = - \partial_1 \chi \hspace{20mm} ,\hspace{20mm} P_\chi = - \partial_1 y.
\end{equation}
From now on we will work in conformal gauge with $h_{_{00}} = -1 $ and $h_{_{11}} = 1$, so that the above canonical transformation becomes equivalent to the following nonlocal map [6,7]:  
\begin{equation}
\partial_\alpha y = -\tilde{B}_{0 \mu} \partial_\alpha x^\mu + {\epsilon_\alpha}^\beta (\tilde{G}_{00} \partial_\beta \chi + \tilde{G}_{0 \mu} \partial_\beta x^\mu).
\end{equation}

\section{$\sigma$-model with Kaluza-Klein Metric (Pole Solution)} 
\subsection{The Model}
In this section we consider a $\sigma$-model with metric of target manifold $M$ as Kaluza-Klein metric with $U(1)$ isometry.   
Hence for its metric we have \cite{K}:
\begin{equation}
ds^2=G_{AB}{dX^A}{dX^B}=g_{\mu \nu} dx^\mu dx^\nu + R^2 (dy + \frac{k A_\mu}{R} dx^\mu )^2 .  
\end{equation}
Here $x^A = \{y,x^\mu\}$, $ A = 0,1,2,3,5$ , and  $\mu = 0,1,2,3 $. The metric given by (7) can also be written as: 
\begin{equation}
G_{AB} =  \left( \begin{array}{ll}
                   g_{\mu \nu} + k^2 A_\mu A_\nu  &  k R A_\mu  \\ 
                   k R A_\mu                &  R^2         \\  
                   \end{array}  \right) . 
\end{equation}
Here we have considered the abelian Kaluza-Klein theory with $A_\mu$ as an electromagnetic gauge field, while $x^\mu$ and $g_{\mu \nu}$ are the coordinates and metric of four  
dimensional space-time respectively. The constant $k$ is introduced in order that $kA_\mu$ becomes dimensionless. 
Also we have taken the extra fifth dimension as a circle with radius $R$ and an angular variable as its coordinate.  
In the definition of Kaluza-Klein metric given in (7) all terms of the five dimensional world-line are dimensionally consistent.    
Notice that the gauge field $A_\mu$ and the metric $g_{\mu \nu}$ are independent of $y$, hence the fifth coordinate is the isometry one.  
Indeed the Kaluza-Klein metric (7) is the metric of a $U(1)$ principal bundle. Actually, in this section we have taken the target manifold of the $\sigma$-model as a $U(1)$ principal bundle.  
According to the formula (1), the action of this model in a conformal gauge with vanishing torsion can be written as: 
\begin{equation}
S=-\frac{1}{4\pi{\alpha^\prime}}\int\!d^2\xi G_{AB}{\partial_\alpha}{X^A}{\partial^\alpha}{X^B} .
\end{equation}
Its variation leads to the following Euler-Lagrange equation:
\begin{equation}
\partial_\alpha(G_{AB}\partial^\alpha X^B) - \frac{1}{2} {\partial_A}G_{BC} \partial_\alpha X^B {\partial^\alpha} X^C =0 .  
\end{equation}
As $G_{AB}$ is independent of the fifth coordinate $y$, therefore $A=5$ case leads to:  
\begin{equation}
{\partial_\alpha} a^\alpha = 0,\hspace{20mm} {a^\alpha} = G_{5B} \partial^\alpha X^B .
\end{equation}
This is the same as the conservation law due to isometry, and if we use the explicit form of Kaluza-Klein metric   
(8), this relation can be interpreted as a continuity relation. As we will see below, it can be obtained via Noether theorem. 
Using the explicit form of the Kaluza-Klein metric (8), the equation of motion (10) for $A=\mu$ components can be written as:  
\begin{equation}
\partial_\alpha(g_{\mu \nu} \partial^\alpha x^\nu) - \frac{1}{2} \partial_\mu g_{\nu \lambda} \partial_\alpha x^\nu \partial^\alpha x^\lambda = \frac{k}{R}{a_\alpha} F_{\mu \nu} \partial^\alpha x^\nu ,
\end{equation}
where $F_{\mu \nu} = \partial_\mu A_\nu - \partial_\nu A_\mu$ is the electromagnetic field intensity.
This is the equation of a two dimensional $\sigma$-model in the presence of an electromagnetic field. 
Let us consider the following solitonic solution for the equation (10):  
\begin{equation}
y = m \sigma + \lambda \tau,\hspace{20mm}     x^\mu = x^\mu(\tau) ,
\end{equation}
with the coordinate $\xi = (\xi^0 , \xi^1)$ of the world sheet as $\xi = (\tau , \sqrt{\alpha^\prime} \sigma)$, where $0 \leq \sigma \leq 2\pi$; $\sqrt{\alpha^\prime}$ appears for dimensional consistency (we are working in natural system of units);  
m is an arbitrary integer and $\lambda$ is a constant. This ansatz has also been proposed in reference \cite{Gibbons}, but there are some important points which deserve much attention. Here we are 
going to deal with $\sigma$-model, hence $h^{\alpha \beta}$ is an intrinsic metric and it differs drastically from the metric induced via embedding of two dimensional world sheet into target manifold $M$.
Therefore, contrary to reference \cite{Gibbons}, there is no contradiction with the choice of $h^{\alpha \beta}$ as a conformal metric.    
Now, by using the ansatz (13) in equations (12) and (11), we get:   
\begin{equation}
a_\tau = k R A_\mu {\dot{x}}^\mu + R^2 \lambda = const,\hspace{40mm}  a_\sigma = \frac{m R^2}{\sqrt{\alpha^\prime}}
\end{equation} 
\begin{equation}
\frac{d}{d\tau}(g_{\mu \nu} {\dot{x}}^\nu) - \frac{1}{2}\partial_\mu g_{\nu \lambda} {\dot{x}}^\nu {\dot{x}}^\lambda = \frac{k}{R}{a_\tau} F_{\mu \nu} {\dot{x}}^\nu .
\end{equation} 
This is the same as the geodesic equation of a charged point particle in the curved space-time with metric $g_{\mu \nu}$, within the presence of an electromagnetic field $A_\mu$.   
Now, if we assume that $A_\mu$ is the gauge field of a monopole located at the origin (consistent with the fact that monopole is a solitonic solution of five dimensional Einstein equation \cite{Gross}), then 
equation (15) can be interpreted as the equation of motion of a point   
charged particle in a given curved space-time within the presense of a magnetic monopole.    
In order to make the subject more clear, we investigate the Hamiltonian of the system. Hence we     
write action (9) in terms of the space-time metric and gauge field as: 
\begin{equation}
S = -\frac{1}{4 \pi \alpha^\prime}\int\!{{d^2}\xi}[{R^2}{\partial_\alpha}{y}{\partial^\alpha}{y} + 2kR{A_\mu}{\partial_\alpha}{y}{\partial^\alpha}{x^\mu} + (g_{\mu \nu} + {k^2}{A_\mu}{A_\nu}){\partial_\alpha}{x^\mu}{\partial^\alpha}{x^\nu}] .
\end{equation}
Using the ansatz (13), we get the following form for the momentum density: 
\begin{equation}
\Pi_\mu = \frac{1}{2\pi \alpha^\prime}[(g_{\mu \nu} + k^2 A_\mu A_\nu) {\dot{x}}^\nu + kR\lambda A_\mu],\hspace{20mm} \Pi_5 = \frac{a_\tau}{2\pi {\alpha^\prime}} .
\end{equation}
Also, substituting the ansatz (13) in the action (16) we get the following action of a point particle by integrating over $\xi_1$:
\begin{equation} 
S = -\frac{1}{2 \sqrt{\alpha^\prime}}\int\!{d\tau} [\frac{m^2 R^2}{\alpha^\prime} - R^2\lambda^2 - 2kR \lambda{ A_\mu}{\dot{x}}^\mu - (g_{\mu \nu} + k^2 A_\mu A_\nu) {\dot{x}}^\mu {\dot{x}}^\nu] .
\end{equation}
Hence, we have 
\begin{equation}
{\dot{x}}^\mu = \sqrt{\alpha^\prime} g^{\mu \nu} (P_\nu - \frac{k a_\tau}{\sqrt{\alpha^\prime} R} A_\nu) ,
\end{equation}
where $P_\nu$ is the momentum of soliton. Finally, for the Hamiltonian of the system we get: 
\begin{equation}
H = \frac{\sqrt{\alpha^\prime}}{2} g^{\mu \nu} (P_\mu - Q_e A_\mu) (P_\nu - Q_e A_\nu) + (\frac{m^2 R^2}{2 \sqrt{\alpha^\prime} \alpha^\prime} + \frac{\sqrt{\alpha^\prime} Q_e^2}{2 k^2} - \frac{Q_e \lambda R}{k}) ,
\end{equation}
where $Q_e$, the charge of soliton, is defined as: 
\begin{equation}
Q_e = \frac{k a_\tau}{\sqrt{\alpha^\prime} R} .
\end{equation}
It is clear from the formula (20) that the Hamiltonian of the system depends on the integer $m$. Also, from equations (15) and (20) we see that the mass of the soliton is $M=\frac{1}{\sqrt{\alpha^\prime}}$. 
On the other hand, this Hamiltonain is bounded from below by  
$|Q_e|$, which is one of the characters of the solitonic solution. Moreover, as expected from the Kaluza-Klein theory, the  fifth momentum $P_5$ is proportional to the charge of soliton:   
\begin{equation}
P_5 = \int_0^{2\pi}\! \Pi_5 d\sigma = \frac{a_\tau}{\alpha^\prime},\hspace{20mm} Q_e = \frac{k \sqrt{\alpha^\prime}}{R} P_5 .
\end{equation}
Therefore, charge quantization follows from the quantization of the fifth momentum as compactness of the fifth coordinate:  
\begin{equation}
P_5 = \frac{n}{R}, \hspace{40mm} Q_e = \frac{k \sqrt{\alpha^\prime}}{R^2} n ,
\end{equation}
where $n$ is an integer number.
Before proceeding with the investigation of the dual model we derive the topological and Noether currents of the model itself. In order  
to obtain the topological charge we consider the following topological current: 
\begin{equation} 
j_{top}^\alpha = \frac{R^2}{k \sqrt{\alpha^\prime}} \epsilon^{\alpha \beta} \partial_\beta y .
\end{equation}
Hence, the topological charge is defined as: 
\begin{equation}
Q_{top} = \int_0^{2\pi}\!j_{top}^0 d\xi^1 = \frac{2\pi m R^2}{k\sqrt{\alpha^\prime}} .
\end{equation}
This is similar to the magnetic charge of the monopole. To see this, we consider     
the Chern number of $U(1)$ principal bundle: 
\begin{equation}
g = \int_c \! d A \stackrel{S T}{\longrightarrow} \frac{R^2}{k\sqrt{\alpha^\prime}}\int\!dy = Q_{top} .
\end{equation} 
Note that in using the Stokes theorem we have assumed that $A$ approaches an exact form in the boundary of region of surface integration.  
Now, using the relations (23) and (25), we obtain the well-known Dirac's quantization: 
\begin{equation}
Q_e . Q_{top} = 2\pi m n .
\end{equation}
In order to obtain Noether current we investigate the symmetry of the action (16). We have already seen that this action has an isometry symmetry, and it is invariant under the transformation $y \longrightarrow y + \varepsilon $ in the adapted coordinate.     
Therefore, according to Noether theorem we obtain the following conserved current:
\begin{equation}
\partial_\alpha J^\alpha = 0, \hspace{40mm} J^\alpha = \frac{k}{\alpha^\prime R}[kRA_\mu \partial^\alpha x^\mu + R^2 \partial^\alpha y]
\end{equation}
which is the same as relation (11). We can obtain the charge $Q_e$ by integrating over $J^0$. 

\subsection{Dual Model}
Using the transformation (4) we obtain the dual model. In writing these transformation relations one should be careful. For example, the metric $G_{AB}$ and torsion potential $B_{AB}$ have chosen to be dimensionless 
in section two, while some terms in the action (16) have the dimension of length. 
Hence, we must take into account the coefficient $\frac{1}{\alpha^\prime}$ for the sake of dimensional consistency. Also, note that in relation (16) isometry is in the fifth 
dimension, hence in relation (2) the index $0$ plays the role of $5$ in relation (16). 
After taking into account all these facts, the coefficients $\tilde{G}_{AB}$ and $\tilde{B}_{AB}$ of dual model can be written as:   
$$
\tilde{G}_{55}=\frac{\alpha^\prime}{R^2},\hspace{20mm}\tilde{G}_{5 \mu}= 0,\hspace{20mm}\tilde{G}_{\mu \nu}= g_{\mu \nu}/{\alpha^\prime},\\
$$
\begin{equation}
\tilde{B}_{5 \mu}=\frac{k}{R} A_\mu,\hspace{40mm} \tilde{B}_{\mu \nu}= 0  .
\end{equation}
Using these data we get the following action for the dual model: 
\begin{equation}
\tilde{S} = -\frac{1}{4\pi}\int\!{d^2}\xi [{\frac{\alpha^\prime}{R^2}}{\partial_\alpha}{\chi}{\partial^\alpha}{\chi} + \frac{1}{\alpha^\prime} g_{\mu \nu}{\partial_\alpha}{x^\mu}{\partial^\alpha}{x^\nu} + 
2\epsilon^{\alpha \beta}\frac{k}{R} A_\mu{\partial_\alpha}{\chi}{\partial_\beta}{x^\mu}] .
\end{equation}
Before proceeding to discuss this action and the results which follow from it, it would be most convenient to compare it with the action (16).       
Indeed the two actions describe the same physical system but their geometries differ drastically from each other. If we use the method of path-space used in \cite{Zaccaria} 
for describing the dynamics of a system consisting of a charged point particle and a magnetic monopole, the antisymmetric term in action (30) can be interpreted as the 
interaction of a two-dimesional field with the magnetic monopole. Indeed, using the Stoke's theorem, this term can be obtained from the following Wess-Zumino term (in path-space):
$$
\frac{-k}{16\pi R} \epsilon^{\alpha \beta \gamma} \int_B F_{\mu \nu} \partial_\alpha \chi \partial_\beta x^\mu \partial_\gamma x^\nu d^ 3\xi,
$$
where $B$ is a three dimensional manifold with world surface as its boundary. Therefore, the antisymmetric Wess-Zumino term in the action (30) can have physical interpretation. Let us   
see what kind of a system can be described by the action (30), if we use the solitonic solution which is discussed in the previous section. To make the subject  more clear, let us consider the following ansatz:      
\begin{equation}
\chi = \tilde{m} \sigma + \tilde{\lambda} \tau,\hspace{40mm} x^\mu = x^\mu(\tau) ,
\end{equation}
where, similar to the relation (13), $\tilde{m}$ is an integer and $\tilde{\lambda}$ is a constant. By substituting the relations (13) and (31) in equation (6) and using the relation (29), we obtain the following relation between $(\tilde{m} , \tilde{\lambda})$ and $(m , \lambda)$ :  
$$
\tilde{m} = \frac{a_\tau}{\sqrt{\alpha^\prime}},\hspace{20mm} \tilde{\lambda} = \frac{R^2 m}{\alpha^\prime \sqrt{\alpha^\prime}} .
$$
Hence, $\chi$ can be written as: 
\begin{equation} 
\chi = \frac{R}{k} Q_e \sigma + \frac{k}{2\pi \alpha^\prime} Q_{top} \tau .
\end{equation}
Now we investigate the topological and Noether currents. As we have described in the previous section and as it is clear from the form of the action (30), we see that the action has a $U(1)$ isometry. Hence the action (30) is invariant under the transformation      
$\chi \longrightarrow \chi + \epsilon$ . Thus, the Noetherian current follows from this symmetry:  
\begin{equation}
\tilde{J}_\alpha = \frac{\alpha^\prime}{R^2} \partial_\alpha \chi + {\epsilon_\alpha}^\beta\frac{k}{R} A_\mu \partial_\beta x^\mu . 
\end{equation}
Hence, the Noetherian charge of solitonic solution (31) is equal to      
\begin{equation}
\tilde{Q}_{_{N}} = \int_0^{2\pi}\tilde{J}_0 d\xi_1 = \frac{k\sqrt{\alpha^\prime}}{R^2} Q_{top}  .
\end{equation}
On the other hand, the topological current of the system (30) can be written as:  
\begin{equation}
{\tilde{j}_{top}}^\alpha = \frac{R}{2\pi \sqrt{\alpha^\prime}} \epsilon^{\alpha \beta} \partial_\beta \chi .
\end{equation}
Therefore, for the topological charge we get
\begin{equation}
\tilde{Q}_{top} = \frac{R^2}{k \sqrt{\alpha^\prime}} Q_e  .
\end{equation}
Comparing the relations (34) and (36) we see that the Buscher's or dual transformation interchanges the electric and magnetic charges. Hence for solitonic solutions of kind (13) or (31),  
T-duality transformation interchanges the Noether charge following from the equation of motion with the topological charge associated to the Bianchi identity; that is it acts as electric-magnetic transformation.   
Also, note that from equations (34) and (36) we get: 
\begin{equation}
{\tilde{Q}}_N.{\tilde{Q}}_{top} = Q_e.Q_{top} 
\end{equation}

\section{$\sigma$-model with Kaluza-Klein Metric with Torsion Term (Dyonic Solution)}
\subsection{The Model}
In this section we consider again the $\sigma$-model with Kaluza-Klein metric with an extra antisymmetric term of the form: 
$$
S = -\frac{1}{4\pi\alpha^\prime} \int\!d^2\xi[R^2 \partial_\alpha y \partial^\alpha y + 2kR {\cal A}_\mu \partial_\alpha y \partial^\alpha x^\mu + (g_{\mu \nu} + k^2{\cal A}_\mu{\cal A}_\nu)\partial_\alpha  x^\mu \partial^\alpha x^\nu      
$$
\begin{equation}
\hspace{20mm} + 2\epsilon^{\alpha \beta} kR{\tilde{\cal A}}_\mu \partial_\alpha y \partial_\beta x^\mu] ,
\end{equation}
where ${\cal A}_\mu$ and ${\tilde{\cal A}}_\mu$ are non-local gauge potentials \cite{Cardoso}. As a result, the electromagnetic intensity field and its dual can be written in terms of these potentials:  
\begin{equation}
\begin{array}{ll}
F_{\mu \nu} = \partial_\mu{\cal A}_\nu - \partial_\nu{\cal A}_\mu, \\
\\
\tilde{F}_{\mu \nu} = \partial_\mu{\tilde{\cal A}}_\nu - \partial_\nu{\tilde{\cal A}}_\mu ,
\end{array}
\end{equation}
where $\tilde{F}_{\mu \nu}=\frac{1}{2}\epsilon_{\mu \nu \lambda \eta} F^{\lambda \eta}$.
It has already been shown in reference \cite{Cardoso} that the non-locality of potentials ${\cal A}_\mu , {\tilde{\cal A}}_\mu$ affects neither the equation of motion nor $F_{\mu \nu}$ and its dual $\tilde{F}_{\mu \nu}$.  
In the action (38), the metric $g_{\mu \nu}$ and potentials ${\cal A}_\mu , {\tilde{\cal A}}_\mu$ are assumed to be independent of $y$. Of course, this assumption is compatible with the non-locality of potentials. Similar to the previous 
section, $y$ is the coordinate of isometry. In general, variation of the action (1) leads to the following equations of motion:     
\begin{equation}
\partial_\alpha[(\delta^{\alpha \beta}G_{AB} + \epsilon^{\alpha \beta} B_{AB}){\partial_\beta X^B}] - \frac{1}{2} \partial_A(\delta^{\alpha \beta} G_{BC} + \epsilon^{\alpha \beta} B_{BC})\partial_\alpha X^B \partial_\beta X^C = 0 .
\end{equation}
Again, since $G_{AB}$ and $B_{AB}$ of the action (38) are independent from the coordinate $y$, we get the following equation for $A=5$:  
\begin{equation}
\partial_\alpha d^\alpha = 0,  \hspace{40mm} d^\alpha = R^2{\partial^\alpha y} + kR{\cal A}_\mu \partial^\alpha x^\mu + \epsilon^{\alpha \beta} kR{\tilde{\cal A}}_\mu \partial_\beta x^\mu , 
\end{equation}
whereas the $A=\mu$ component of equation (40) takes the form:
$$
\partial_\alpha(g_{\mu \nu}\partial^\alpha x^\nu) - \frac{1}{2}\partial_\mu{g_{\nu \lambda}} \partial_\alpha x^\nu \partial^\alpha x^\lambda = \frac{k}{R}d^\alpha F_{\mu \nu}\partial_\alpha x^\nu + kR \epsilon^{\alpha \beta}\tilde{F}_{\mu \nu} \partial_\alpha y \partial_\beta x^\nu
$$
\begin{equation}
\hspace{60mm}+\frac{k^2}{2}\epsilon^{\alpha \beta}({\cal A}_\mu \tilde{F}_{\nu \lambda} + 2{\tilde{\cal A}_\nu}F_{\mu \lambda})\partial_\alpha x^\nu \partial_\beta x^\lambda .
\end{equation}
Now, similar to the previous section, we consider the following solitonic ansatz: 
\begin{equation}
y = m \sigma, \hspace{40mm} x^\mu = x^\mu(\tau) .
\end{equation}
In this case, the equation of motion reduces to: 
$$
d_\tau = kR{\cal A}_\mu {\dot{x}}^\mu = const,
$$
and
\begin{equation}
\frac{d}{d\tau}(g_{\mu \nu} {\dot{x}}^\nu) - \frac{1}{2}\partial_\mu g_{\nu \lambda} {\dot{x}}^\nu{\dot{x}}^\lambda = (\frac{k d_\tau}{R}F_{\mu \nu} + \frac{mkR}{\sqrt{\alpha^\prime}}\tilde{F}_{\mu \nu}){\dot{x}}^\nu ,
\end{equation}
which can be interpreted as equation of motion of dyon in a space-time with metric $g_{\mu \nu}$ in the presence of electromagnetic field $F_{\mu \nu}$. Similar 
to the previous section, to see how the solitonic solution can appear as dyon, we consider the Hamiltonian of the system: 
\begin{equation}
H = \frac{\sqrt{\alpha^\prime}}{2}g^{\mu \nu}(P_\mu - Q_{_{E}}{\cal A}_\mu - Q_{_{M}}{\tilde{\cal A}}_\mu)(P_\nu - Q_{_{E}}{\cal A}_\nu - Q_{_{M}}{\tilde{\cal A}}_\nu) + \frac{\sqrt{\alpha^\prime}}{2k^2}({Q_{_{E}}}^2 + {Q_{_{M}}}^2) ,
\end{equation}
where the electric and magnetic charges of dyon are defined as: 
\begin{equation}
Q_{_{E}} = \frac{kd_\tau}{\sqrt{\alpha^\prime}R} \hspace{20mm},\hspace{20mm} Q_{_{M}} = \frac{kRm}{\alpha^\prime} .
\end{equation}
It is clear from the equation of motion (44) that the dyon has the mass $M = \frac{1}{\sqrt{\alpha^\prime}}$. It is also obvious from the Hamiltonian (45) that the solitonic solution saturates the Bogomolnyi limit:  
\begin{equation}
{\cal E} \geq \sqrt{{Q_{_{E}}}^2 + {Q_{_{M}}}^2} .
\end{equation}
We can define the fifth momentum using the relation (38) and analogous to the previous section, we can obtain a relation between fifth momentum and electric charge of dyon. Now we proceed to investigate the solitonic solution in dual model and give its physical interpretation.    
\subsection{Dual Model}
In order to obtain the dual model we use again the Buscher's transformation (4); the coefficients of the metric and the torsion terms are chosen to be dimensionally consistent. Hence, we have:      
$$
\tilde{G}_{55}=\frac{\alpha^\prime}{R^2},\hspace{20mm}\tilde{G}_{5 \mu}= \frac{k}{R}{\tilde{\cal A}}_\mu,\hspace{10mm}\tilde{G}_{\mu \nu}= (g_{\mu \nu} + k^2{\tilde{\cal A}}_\mu {\tilde{\cal A}}_\nu)/{\alpha^\prime},
$$
\begin{equation}
\tilde{B}_{5 \mu}=\frac{k}{R}{\cal A}_\mu,\hspace{40mm} \tilde{B}_{\mu \nu}= -\frac{k^2}{\alpha^\prime}({\cal A}_\mu {\tilde{\cal A}}_\nu - {\cal A}_\nu {\tilde{\cal A}}_\mu) . 
\end{equation}
Therefore, the action of the dual model can be written as: 
$$
\tilde{S} = -\frac{1}{4\pi}\int\!d^2\xi[\frac{\alpha^\prime}{R^2}{\partial_\alpha \chi}{\partial^\alpha \chi} + \frac{2k}{R}{\tilde{\cal A}_\mu}{\partial_\alpha \chi}{\partial^\alpha} x^\mu + \frac{1}{\alpha^\prime}(g_{\mu \nu} + 
k^2{\tilde{\cal A}_\mu}{\tilde{\cal A}_\nu}){\partial_\alpha x^\mu}{\partial^\alpha x^\nu}
$$
\begin{equation}
\hspace{35mm} +\frac{2k}{R} \epsilon^{\alpha \beta}{\cal A}_\mu{\partial_\alpha \chi}{\partial_\beta x^\mu} - \frac{k^2}{\alpha^\prime} \epsilon^{\alpha \beta}({\cal A}_\mu{\tilde{\cal A}_\nu} - {\cal A}_\nu{\tilde{\cal A}_\mu}){\partial_\alpha x^\mu}{\partial_\beta x^\nu}] .
\end{equation}
The equation of motion is same as (40) provided that we replace the isometry coordinate $y$, the metric $G_{AB}$ and torsion $B_{AB}$ by the dual isometry coordinate $\chi$, dual metric $\tilde{G}_{AB}$ and dual torsion $\tilde{B}_{AB}$, respectively.    
Thus, for the following solitonic solution similar to (43): 
\begin{equation}
\chi = \tilde{m}\sigma, \hspace{60mm}x^\mu = x^\mu(\tau) ,
\end{equation}
where $\tilde{m}$ is an integer, the equations of motion become: 
$$
{\tilde{d}}_\tau = \frac{k}{R}{\tilde{\cal A}}_\mu {\dot{x}}^\mu = const,
$$
\begin{equation}
\frac{d}{d\tau}(g_{\mu \nu}{\dot{x}}^\nu) - \frac{1}{2}\partial_\mu g_{\nu \lambda}{\dot{x}}^\nu{\dot{x}}^\lambda = (kR\tilde{d}_\tau\tilde{F}_{\mu \nu} + \frac{\tilde{m}k\sqrt{\alpha^\prime}}{R} F_{\mu \nu}){\dot{x}}^\nu .
\end{equation}
These equations indicate that even the solitonic solution of the dual model has a dyonic behaviour. To find the exact value of the electric and magnetic charge, and in order to further investigate the behaviour of solitons, we consider the Hamiltonian of the system: 
\begin{equation}
\tilde{H} = \frac{\sqrt{\alpha^\prime}}{2}g^{\mu \nu}(P_\mu - \tilde{Q}_{_{E}}{\cal A}_\mu - \tilde{Q}_{_{M}}\tilde{\cal A}_\mu)(P_\nu - \tilde{Q}_{_{E}}{\cal A}_\nu - \tilde{Q}_{_{M}}\tilde{\cal A}_\nu) + \frac{\sqrt{\alpha^\prime}}{2k^2}({\tilde{Q}_{_{E}}}^2 + 
{\tilde{Q}_{_{M}}}^2), 
\end{equation}
with the electric charge $\tilde{Q}_{_{E}}$ and the magnetic charge $\tilde{Q}_{_{M}}$ of dyon as:     
\begin{equation}
\tilde{Q}_{_{E}} = \frac{k\tilde{m}}{R},\hspace{40mm} \tilde{Q}_{_{M}} = \frac{kR\tilde{d}_\tau}{\sqrt{\alpha^\prime}} .
\end{equation}
Again, from the dual Hamiltonian (52), it is clear that the solitonic solution (50) saturates Bogomolnyi limit:   
\begin{equation}
\tilde{\cal E} \geq \sqrt{{\tilde{Q}_{_{E}}}^2 + {\tilde{Q}_{_{M}}}^2}  .
\end{equation}
Now in order to get the relation between the model and its dual, we compare the equations (44) and (45) with their duals (51) and (52). We see that duality transformation (48) is equivalent with the following one: 
\begin{equation}
{\cal A}_\mu \longleftrightarrow \tilde{\cal A}_\mu,\hspace{40mm} R \longleftrightarrow \frac{\alpha^\prime}{R}
\end{equation}
or consequently: 
\begin{equation}
m \longleftrightarrow \tilde{m},\hspace{20mm} Q_{_{E}} \longleftrightarrow \tilde{Q}_{_{M}},\hspace{20mm} Q_{_{M}}\longleftrightarrow \tilde{Q}_{_{E}} .
\end{equation}
Now, using the relation (6) for the solitonic solutions (50) and (43), we get the following relations between the charges:   
\begin{equation}
\tilde{Q}_{_{M}} = Q_{_{M}}, \hspace{60mm} \tilde{Q}_{_{E}} = Q_{_{E}}   .
\end{equation} 
Hence, duality transformation is the same as $Q_{_{E}} \longleftrightarrow Q_{_{M}}$, that is the electric-magnetic duality transformation.

\section{Conclusion}
We saw that the torsionless $\sigma$-model with Kaluza-Klein metric for some special solitonic ansatz reduced to a system  
which described the motion of a charged particle in a given curved background manifold, where for a strong coupling ($T=\frac{1}{2\pi\alpha^\prime}$) the mass of this soliton was $M=\frac{1}{\sqrt{\alpha^\prime}}$.  
Also, by investigating its dual model we have shown that usual abelian dual transformation 
interchanges the Noether charge obtained from the equation of motion with topological charge associated to the Bianchi identity. We also obtained these results       
for the $\sigma$-model with torsion term using the non-local gauge potential. Again it has been shown that for some solitonic ansatz this model reduced to a system  
describing the motion of a dyon in a given curved background manifold, where similary for the mass of dyon we have $M=\frac{1}{\sqrt{\alpha^\prime}}$. Finally, we saw that the abelian  
duality leads to the interchange of electric and magnetic charge of dyon. In other words, for this special ansatz, T-duality can be interpreted as electric-magnetic duality.

\vskip8pt\noindent

{\large{\bf Acknowledgment}}
We wish to thank  Dr. S.K.A. Seyed Yagoobi for carefully reading the 
article and for his constructive comments.

\end{document}